# Estimation & Recognition under Perspective of Random-Fuzzy Dual Interpretation of Unknown Quantity: with Demonstration of IMM Filter

Wei Mei, Yunfeng Xu, Limin Liu

*Abstract*—This paper is to consider the problems of estimation and recognition from the perspective of sigma-max inference (probability-possibility inference), with a focus on discovering whether some of the unknown quantities involved could be more faithfully modeled as fuzzy uncertainty. Two related key issues are addressed: 1) the random-fuzzy dual interpretation of unknown quantity being estimated; 2) the principle of selecting sigma-max operator for practical problems, such as estimation and recognition. Our perspective, conceived from definitions of randomness and fuzziness, is that continuous unknown quantity involved in estimation with inaccurate prior should be more appropriately modeled as randomness and handled by sigma inference; whereas discrete unknown quantity involved in recognition with insufficient (and inaccurate) prior could be better modeled as fuzziness and handled by max inference. The philosophy was demonstrated by an updated version of the well-known interacting multiple model (IMM) filter, for which the jump Markov system is reformulated as a hybrid uncertainty system, with continuous state evolution modeled as usual as model-conditioned stochastic system and discrete mode transition modeled as fuzzy system by a possibility (instead of probability) transition matrix, and hypotheses mixing is conducted by using the operation of "max" instead of "sigma". For our example of maneuvering target tracking using simulated data from both a short-range fire control radar and a long-range surveillance radar, the updated IMM filter shows significant improvement over the classic IMM filter, due to its peculiarity of hard decision of system model and a faster response to the transition of discrete mode.

## I. Introduction

Many tasks of signal/information processing involve the handling of uncertainty since the available prior information are often inaccurate or contaminated by noise. The inaccuracy or noise is usually regarded as randomness and is expressed, as recommended by Joint Committee for Guides in Metrology (JCGM) [1], by using probability density function. Therefore, for typical tasks of information processing, such as estimation or recognition, the unknown quantity to be estimated or recognized would usually as well be modeled as random uncertainty by employing probability theory - the dominant theory for handling uncertainty in the community of information sciences. In fact, probability theory and the derivative subjects of statistics and stochastic process so far remain the most popular and rigorous tool set for the problems of estimation and recognition [2,3]. On the other hand, it is well realized that uncertainty may still arise for the task of recognition of unknown quantity (pattern) when the available prior information is accurate but insufficient. The insufficiency of the prior information lies in that the pattern or concept, which the unknown quantity will be classified into, themselves have uncertainty (e.g., fuzzy uncertainty) because the intensions of the them are overlapped [4]. Here intension indicates the internal content (abstracted feature) of a concept that constitutes its formal definition [4,5]. In other words, by insufficiency we mean the feature that can be extracted from the prior information cannot be used to recognize the unknown quantity without ambiguity. As a classical example, let us consider the fuzzy concept of Young. Given the exact age of 42 of a man, we will be hesitated to recognize (the age group of) him into Young or Middle Age. In other words, age of 42 cannot be recognized as Young or Middle Age without ambiguity. Here the unknown quantity is age group, and the uncertainty occurred in recognizing the unknown quantity is known as fuzzy uncertainty [4]. The information of age is not sufficient for accurately recognizing the unknown quantity of age group, and Young and Middle Age need to be regarded as fuzzy concept because the intensions of the them are overlapped.

Estimation and (pattern) recognition play important roles in signal/ information processing, automatic control, communication, machine intelligence, and a diversity of other fields. Estimation is the process of estimating the value of a quantity of interest from indirect, inaccurate, and uncertain observations [2]. Pattern recognition consists of identifying a given pattern which is known to belong to one of a finite set of classes [3,6]. Follow these definitions and to our understanding, estimation and recognition have two basic differences. First, pattern recognition can be viewed as the process of selecting one out of a set of *discrete* alternatives, whereas estimation cares about the selection of a point from a *continuous* space — the best estimate [2]. Second, pattern recognition usually comes along with the process of feature extraction, which finds the appropriate features for representing the input patterns [3]. The common

Wei Mei (meiwei@sina.com) and Limin Liu (lk0256@163.com) are with the Electronic Engineering Department, Army Engineering University, Shijiazhuang, 050003, P.R. China. Yunfeng Xu (hbkd_xyf@hebust.edu.cn) is working with both Army Engineering University and Hebei University of Science and Technology, Shijiazhuang 050018, China.

place of estimation and recognition lies in that both problems are devoted to infer the unknown quantity from available inaccurate measurements. For pattern recognition, the unknown quantity is *discrete* and represents the categories of the pattern. For estimation, the unknown quantity is usually *continuous* and represents the value of parameter or state being estimated. A parameter is a time-invariant quantity and the state of a dynamic system would evolve in time, hence estimation could be categorized into parameter estimation and state estimation [2,7].

Here is then a natural question how we should interpret and model the uncertainty that is arisen in the problems such as estimation and recognition of unknown quantity given inaccurate or insufficient prior knowledge. We will present our perspective on this after giving more introduction to the above-mentioned concepts of randomness and fuzziness, and the related uncertainty theories of probability and possibility. Randomness and fuzziness are widely acknowledged as two kinds of fundamental uncertainties of this world, yet clear and well-known definitions for them had been lacking. Recently, randomness and fuzziness are defined in [4,8-10] as below:

*Randomness is the occurrence uncertainty of the either-or outcome of a causal experiment, characterized by the lack of predictability in mutually exclusive outcomes.*

*Fuzziness is the classification uncertainty of the both-and outcome of a cognition experiment, characterized by the lack of clear boundary between non-exclusive outcomes.*

In fact, the topic on fuzzy uncertainty is not new and the theory of fuzzy sets has been widely investigated since its origin in 1965 by Zadeh [11]. Nevertheless, applications of fuzzy sets are basically limited to fuzzy inference system [12,13] and later in fuzzy modeling and control of nonlinear systems [14,15]. In recent years, possibility theory based on the well-known axiom of "maxitivity" was recognized as an alternative method for modeling fuzzy uncertainty [50-56], and is gradually growing to exhibit itself as a potential foundation for fuzzy sets [9,16]. It is now clear that membership function of fuzzy sets can be recognized as likelihood function of possibility (instead of regarding membership function as possibility by Zadeh) [9,16], and composition of fuzzy relations is equal to composition of conditional possibilities [9]. Possibility theory is comparable to probability theory because they are both distribution-based and describe uncertainty with numbers in the unit interval [0, 1]. The difference lies in that probability satisfies the key axiom of "additivity", we hence follow the appellation of [10] to name probability inference as sigma inference and possibility inference as max inference.

Overall, the role of fuzzy sets and possibility theory played in information sciences is far from matching that of probability theory, due to the lack of consensus on the issues pertinent to the foundation of fuzzy sets and possibility theory [4,17-19]. From our perspective, only by making clear the essential difference between randomness and fuzziness can we build up a solid theory of fuzzy uncertainty that is supposed to be different from probability theory. From the clearly-defined concepts of randomness and fuzziness, the latest literature of [4] managed to induce sigma system and max system, respectively. The general conclusion is: probability theory obeys the key axiom of "additivity" because random outcomes are mutual exclusive; whereas possibility theory is marked by the "maxitivity" axiom since fuzzy outcomes are non-exclusive [4]. For many practical problems of information processing, it is natural for us to imagine that random uncertainty may often co-exist with fuzzy uncertainty. As discussed above, e.g., given inaccurate and insufficient prior information, the recognition of unknown pattern may involve both randomness and fuzziness. In [10], a mechanism called sigma-max hybrid uncertainty inference (in short, sigma-max inference) was derived by developing the separate uncertainty systems of probability and possibility into an integrated sigma-max system. This mechanism can cope with randomness and fuzziness jointly and achieve a direct fusion of heterogeneous information modeled by probability or possibility. We hereafter in this work use sigma-max inference to refer to either the sigma-max hybrid uncertainty inference or sigma inference & max inference or both, whenever the reference in the context could be understood without confusion.

The question presented above can now be simply put as: how we should interpret and model the unknown quantity, being estimated or recognized, by means of randomness or fuzziness. Our perspective in this work is that 1) unknown quantity could be dually interpreted as randomness and fuzziness; 2) continuous unknown quantity involved in estimation with inaccurate prior should be more appropriately modeled as randomness, hence should be handled by using sigma inference; 3) discrete unknown quantity involved in recognition with insufficient (and inaccurate) prior could be better modeled as fuzziness, hence should be handled by using max inference (or sigma-max inference). Our perspective is conceived from definitions of randomness and fuzziness, and the functions built in with the "sigma" and "max" operators.

In the rest of this paper, we will elaborate and demonstrate our perspective on the interpretation and modeling of unknown quantity. In section 2, some technique background of sigma-max inference is reviewed, which will be used in section 5 for the development of an update version of the well-known interacting multiple model (IMM) filter [20,21] - hybrid inference IMM (HIMM) filter. In section 3, the problems of estimation & recognition are presented in forms of both sigma inference and max inference, with the unknown quantities involved modeled as random uncertainty and fuzzy uncertainty, respectively. Section 4 investigates the modeling principle of unknown quantity in terms of randomness and fuzziness, and the choice of sigma operator and max operator for the problems of estimation and recognition. In section 6, the HIMM filter developed in section 5 is compared with the classic IMM filter to demonstrate our perspective.

## II. SIGMA-MAX INFERENCE

This section reviews some fundamental aspects of possibility theory and the sigma-max inference. For more detail, the reader may refer to [4,10,22,23].

### A. Random/Fuzzy Variables

Definitions below of random variable and fuzzy variable are excerpted from [4,10].

**Definition 2.1.** A *random variable X* is a variable whose value $x_i$ is subject to variations due to random uncertainty. A random variable can take on a set of possible values in a random sample space $\Omega$, or its generated event space $F \subseteq 2^\Omega$.

**Definition 2.2.** A *fuzzy variable X* is a variable whose value $x_i$ is subject to variations due to fuzzy uncertainty. A fuzzy variable can take on a set of possible values in a fuzzy sample space $\Psi$, or its generated event space $\Sigma \subseteq 2^\Psi$.

*Remark*: A random variable should be handled by sigma inference and fuzzy variable by max inference [4]. Though events in $F \subseteq 2^\Omega$ and $\Sigma \subseteq 2^\Psi$ are both not mutually exclusive, the structures of random event space $F \subseteq 2^\Omega$ and fuzzy event space $\Sigma \subseteq 2^\Psi$ are not the same because their corresponding sample spaces $\Omega$ and $\Psi$ are defined differently [4].

### B. Max Inference

#### 1) Possibility Measure

**Definition 2.3.** *Possibility* (intuitive definition) $\pi_X(x_i)$ is the measure of the both-and fuzziness, which is the confidence of classifying an object $X$ into concept $x_i$. Possibility $\pi_X(x_i)$ of the outcome $x_i$ can be numerically described by the compatibility between a fuzzy variable $X$ and its prospective outcome $x_i$, which can be defined as [4,8]

$$\pi_X(x_i) = \text{comp}(X, x_i) = \text{degree}(f_X \subseteq f_{x_i}) = \frac{|f_X \cap f_{x_i}|}{|f_X|}, \quad (1)$$

where "comp" means compatibility; $f_X$ and $f_{x_i}$ are sets of intensions of the fuzzy variable $X$ and the concept $x_i$, respectively; and $|\cdot|$ is the cardinality or measure.

Note that $\pi_X(x_i)$ is usually and hereafter denoted as $\pi(x_i)$ for short. The interpretation of intuitive possibility can be visualized by Fig. 1, where the frequency interpretation of probability is also presented for contrast. Note that intension of a concept would be obtained through the process of feature extraction, and subsethood measure refers to $\text{degree}(f_X \subseteq f_{x_j})$ as used in (1) [4,23]. The lack of clear boundary between non-exclusive outcomes as mentioned in the definition of fuzziness is caused by the overlap of the intensions of non-exclusive outcomes.

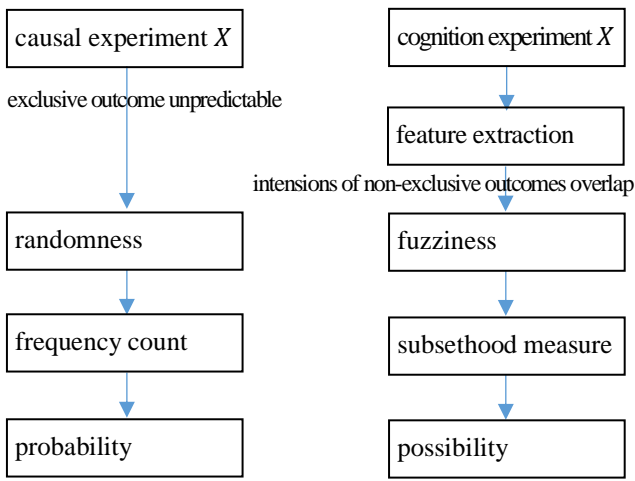

Fig. 1 Randomness & fuzziness and their measures.

**Definition 2.4.** *Possibility* (axiomatic definition) on universal set $\Psi$ is defined as a mapping $\pi: 2^\Psi \to [0,1]$ such that [22,23],

Axiom 1. $\pi(\phi) = 0$ for empty set $\phi$.
Axiom 2. $\pi(\Psi) = 1$.
Axiom 3. $\forall A, B \subseteq \Psi, \pi(A \cup B) = \max\{\pi(A), \pi(B)\}$.

The possibility defined above is usually called normalized or normal possibility and Axiom 3 is the well-known maxitivity axiom, which makes it distinguished from probability. By Axiom 2 and 3 we can derive

$$\pi(\Psi) = \max_{x_i} \pi(x_i) = 1. \quad (2)$$

Eq. (2) indicates that at least one of the elements of $\Psi$ should be fully possible, i.e. $\exists x_i$, such that $\pi(x_i) = 1$. In contrast, probability obeys to the sigma normalization below:

$$\Sigma_{i=1}^N p(x_i) = 1. \quad (3)$$

Suppose $\pi(x_i y_j)$ is the joint possibility distribution of fuzzy variables $X$ and $Y$, then conditional possibility $\pi(y_j|x_i)$ is defined as below [10,22,23]

$$\pi(x_i y_j) = \pi(y_j|x_i)\pi(x_i), \quad (4)$$

where

$$\pi(x_i) = \max_{y_j} \pi(x_i y_j), \; \pi(y_j) = \max_{x_i} \pi(x_i y_j). \quad (5)$$

#### 2) Composition of Fuzzy Relations

Suppose $\pi(y_j|x_i)$ and $\pi(z_k|y_j)$ represent fuzzy relations from $X$ to $Y$ and from $Y$ to $Z$, respectively, then fuzzy relation $\pi(z_k|x_i)$ from $X$ to $Z$ can be given by [9,10]

$$\pi(z_k|x_i) = \max_{y_l} \pi(z_k, y_l|x_i) = \max_{y_l} \pi(z_k|y_l)\pi(y_l|x_i). \quad (6)$$

Composition of stochastic relations as in (7) has a similar structure to its counterpart of (6) except that max operator is replaced with sigma operator.

$$p(z_k|x_i) = \sum_{y_l} p(z_k|y_l)p(y_l|x_i) \quad (7)$$

#### 3) Possibility Update

From (4) and (5) we can derive a possibility update equation [10,24] in a form parallel to Bayesian inference as

$$\pi(x_i|y_j) = \frac{\pi(x_i)\pi(y_j|x_i)}{\pi(y_j)} = \frac{\pi(x_i)\pi(y_j|x_i)}{\max_{x_k} \pi(x_k)\pi(y_j|x_k)}, \quad (8)$$

where $\pi(x_i|y_j)$ is posteriori possibility, $\pi(x_i)$ is priori possibility, and $\pi(y_j|x_i)$ is possibility likelihood of $x_i$.

## C. Sigma-Max Inference

### 1) Hybrid Distribution of Probability/Possibility

**Definition 2.5.** *Hybrid distribution of probability and possibility* is the joint distribution of a random variable $X$ and a fuzzy variable $Y$, which is denoted by [10]

$$p\pi(x_i y_j), \text{ or } \pi p(y_j x_i), \qquad (9)$$

where $\pi$ and $p$ indicate possibility and probability, respectively, and $p\pi(x_i y_j)$ should satisfy normalization requirements (10) and/or (11) below, which are order-dependent with respective to $x_i$ and $y_j$.

$$\sum_{x_i} \max_{y_j} p\pi(x_i y_j) = 1 \qquad (10)$$

$$\max_{y_j} \sum_{x_i} p\pi(x_i y_j) = 1 \qquad (11)$$

Conditional probability $p(x_i|y_j)$ and conditional possibility $\pi(y_j|x_i)$ can be defined as in (12) and (13) below.

$$p\pi(x_i y_j) = p(x_i|y_j)\pi(y_j) \qquad (12)$$

$$p\pi(x_i y_j) = \pi(y_j|x_i)p(x_i) \qquad (13)$$

where $p(x_i|y_j)$ denotes the probability of random variable $X$ being $x_i$ given fuzzy variable $Y$ being $y_j$, and $\pi(y_j|x_i)$ is the possibility of $y_j$ conditioned on $x_i$.

For practical applications, prior knowledge is usually given in forms of marginal distribution and conditional distribution, by which a hybrid distribution can be constructed and would meet the demand of (10) or (11), but not both in general [10]. Suppose $p\pi(x_i y_j)$ is the hybrid distribution generated by $p(x_i|y_j)$ and $\pi(y_j)$ as in (12). Then the generated $p\pi(x_i y_j)$ can be re-expanded as

$$p\pi(x_i y_j) = \pi(y_j|x_i)p^+(x_i) = \beta\pi(y_j|x_i)p(x_i) \qquad (14)$$

where $p^+(x_i)$, as defined in (15), is named as induced distribution, which can be easily verified not to be a probability distribution in general case [10]. And $\beta$ is a normalization factor such that $p(x_i)$ defined in (16) is a probability distribution.

$$p^+(x_i) \triangleq \max_{y_j} p(x_i|y_j)\pi(y_j) \qquad (15)$$

$$p(x_i) \triangleq \frac{1}{\beta} \max_{y_j} p(x_i|y_j)\pi(y_j) \qquad (16)$$

Equations (12) and (14) provide a link between two groups of marginal/conditional distributions, i.e., $p(x_i|y_j) / \pi(y_j)$ and $\pi(y_j|x_i)/p^+(x_i)$. Such a connection holds even though $p^+(x_i)$ is in general not a probability distribution. Similarly, hybrid distribution $p\pi(x_i y_j)$ can be generated by $\pi(y_j|x_i)/p(x_i)$ and re-expanded as $p(x_i|y_j)\pi^+(y_j)$, where the induced distribution $\pi^+(y_j)$ is generally not a possibility distribution.

### 2) Composition of Heterogeneous Relations

Suppose $X$ is a random variable, and $Y$ and $Z$ are fuzzy variables. By applying (16), the induced random relation $p^+(x_i|y_j)$ or $p(x_i|y_j)$ from $Y$ to $X$ can be calculated by [10]

$$p(x_i|y_j) = \frac{1}{\beta}p^+(x_i|y_j) = \frac{1}{\beta}\max_{z_k} p(x_i|z_k)\pi(z_k|y_j), \qquad (17)$$

where $x_i$ and $y_j$ are assumed to be stochastically independent given $z_k$, and $\beta$ is a normalization factor such that $p(x_i|y_j)$ is a probability distribution.

The induced $p^+(x_i|y_j)$ is a direct, hence accurate, fusion of the prior knowledge of $p(x_i|z_k)$ and $\pi(z_k|y_j)$. By a normalization factor of $\beta$, we can have $p(x_i|y_j)$, the probability version of $p^+(x_i|y_j)$. Be aware both $p^+(x_i|y_j)$ and $p(x_i|y_j)$ are many-to-many mappings, which can be either distributions over different $x_i$s conditioned on a certain value of $y_j$, or likelihood functions of different $y_j$s given a certain value of $x_i$. In the former case, it would be necessary to transform $p^+(x_i|y_j)$ into $p(x_i|y_j)$. In the latter case, such a conversion would potentially introduce normalization bias and is unnecessary since $\beta$ in general has different values for different $y_j$.

Similarly, an induced fuzzy relation, denoted by $\pi^+(y_j|x_i)$ or $\pi(y_j|x_i)$, can be computed by [10]

$$\pi(y_j|x_i) = \frac{1}{\beta}\pi^+(y_j|x_i) = \frac{1}{\beta}\sum_{z_k} \pi(y_j|z_k)p(z_k|x_i) \qquad (18)$$

where $y_j$ and $x_i$ are assumed to be possibilistically independent given $z_k$, and $\beta$ is a normalization factor such that $\pi(y_j|x_i)$ is a possibility distribution. Similarly, the application of (18) should make clear whether we need a likelihood expansion of $\pi^+(y_j|x_i)$ over $x_i$s or a distribution of $\pi(y_j|x_i)$ over $y_j$s.

### 3) Uncertainty Update with Heterogeneous Information

By (12), (14) and (16), we can derive [10]

$$\pi(y_j|x_i) = \frac{p(x_i|y_j)\pi(y_j)}{\max_{y_l} p(x_i|y_l)\pi(y_l)} \qquad (19)$$

which is the possibility update equation with prior possibility $\pi(y_j)$ and probability likelihood $p(x_i|y_j)$. Similarly, we have [10]

$$p(x_i|y_j) = \frac{\pi(y_j|x_i)p(x_i)}{\sum_{x_k} \pi(y_j|x_k)p(x_k)} \qquad (20)$$

which is the probability update equation with prior probability $p(x_i)$ and possibility likelihood $\pi(y_j|x_i)$.

## III. ESTIMATION & RECOGNITION UNDER PERSPECTIVE OF SIGMA-MAX INFERENCE

In the following, after giving definitions for unknown quantity and the related, we respectively review the problems of estimation and recognition in form of sigma inference, with the unknown quantities involved modeled as random uncertainty. We then reformulate them by using max inference, with the involved unknown quantities interpreted as fuzzy uncertainty. The philosophy for the interpretation of unknown quantity is left to section 4.

### A. Unknown Quantity and Unknown Uncertainty

**Definition 3.1.** *Unknown quantity* is a quantity that is unknown, or remains to be known, by the subject.

*Remark*: An unknown quantity can be a static unknown quantity, which remains fixed or unchanged over time; or a dynamic unknown quantity, which may change over time.

**Definition 3.2.** *Unknown uncertainty* is the uncertainty carried by an unknown quantity that is assumed to take on $N$ possible outcomes $Y = \{x_1, x_2, \ldots, x_N\}$.

**Definition 3.3.** *Unknown uncertain variable X* is a variable whose value $x_i$ is subject to variation due to unknown uncertainty. An unknown uncertain variable (in short, uncertain variable) can take on a set of possible values from $Y = \{x_1, x_2, \ldots, x_N\}$, or its generated power set $\Xi \subseteq 2^Y$.

*Remark*: Uncertain variable can be used to refer to unknown quantity once its possible values are assumed, and the set of $Y$ can be extended into a set of real number. Uncertain variable can be modeled as either random variable or fuzzy variable depending on the available prior information.

### B. State Estimation in Form of Sigma Inference

Consider the following system

$$x_k = f_{k-1}(x_{k-1}) + Gw_k \tag{21}$$

$$z_k = h_k(x_k) + v_k \tag{22}$$

where transition function $f_{k-1}: \mathbb{R}^{d_x} \to \mathbb{R}^{d_x}$ models the evolution of continuous state vector $x_k$ at time $k$ as a first-order Markov process, with modeling error (uncertainty) represented by process noise $w_k \sim \mathcal{N}(0, Q_k)$. Observation function $h_k: \mathbb{R}^{d_x} \to \mathbb{R}^{d_z}$ models the relationship between the state $x_k$ and the kinematic measurement $z_k$ with measurement error (noise) denoted by $v_k \sim \mathcal{N}(0, R_k)$.

Note that (21) and (22) effectively define two probability distributions, transition density $p(x_k|x_{k-1})$ and measurement density $p(z_k|x_k)$, respectively [25]. Under the perspective of Bayesian theory, state estimation is the process of evolving the posterior probability density $p(x_k|z_{1:k})$ given its prior probability density $p(x_{k-1}|z_{1:k-1})$ and the currently acquired $z_k$, where $z_{1:k} = \{z_1, z_2, \ldots, z_k\}$ is the measurement series up to time $k$. It is usually presented as a two-step procedure as below [25-27]:

*State prediction.*

$$p(x_k|z_{1:k-1}) = \int p(x_k|x_{k-1})p(x_{k-1}|z_{1:k-1})dx_{k-1} \tag{23}$$

*State update.*

$$p(x_k|z_{1:k}) = \frac{p(z_k|x_k)p(x_k|z_{1:k-1})}{\int p(z_k|x_k)p(x_k|z_{1:k-1})dx_k} \tag{24}$$

where the denominator of the right side of (24) functions as a normalization factor.

The mean $\hat{x}_{k|k}$ and the covariance $\hat{\Sigma}_{k|k}$ of $p(x_k|z_{1:k})$ can be computed as the point estimate of the state by

$$\hat{x}_{k|k} \triangleq E[x_k|z_{1:k}] = \int x_k p(x_k|z_{1:k}) dx_k \tag{25}$$

$$\hat{\Sigma}_{k|k} \triangleq E[(x_k - \hat{x}_{k|k})(..)^T | z_{1:k}] \tag{26}$$

Under assumptions of linear-Gaussian system with $f_{k-1}$ and $h_k$ respectively replaced by linear transition matrix $F$ and linear observation matrix $H$, the Kalman filter version of (27)~(32), which is the optimal estimation in term of minimum mean square error (MMSE), can be derived [2,26,27]:

*State prediction/prediction covariance.*

$$\hat{x}_{k|k-1} = F\hat{x}_{k-1|k-1} \tag{27}$$

$$\hat{\Sigma}_{k|k-1} = F\hat{\Sigma}_{k-1|k-1}F^T + GQ_k G^T \tag{28}$$

*State/covariance update.*

$$\hat{x}_{k|k} = \hat{x}_{k|k-1} + W_k(z_k - H\hat{x}_{k|k-1}) \tag{29}$$

$$\hat{\Sigma}_{k|k} = \hat{\Sigma}_{k|k-1} - W_k S_k W_k^T \tag{30}$$

where filter gain $W_k$ and innovation covariance $S_k$ are:

$$S_k = H\hat{\Sigma}_{k|k-1}H^T + R_k \tag{31}$$

$$W_k = \hat{\Sigma}_{k|k-1} H^T S_k^{-1} \tag{32}$$

### C. State Estimation in Form of Max Inference

Now consider the case that process noise $w_k$ and measurement noise $v_k$ are modeled by Gaussian possibility function [25,28]:

$$\bar{\mathcal{N}}(x; \mu, \Sigma) = \exp\left(-\frac{1}{2}(x-\mu)^T \Sigma^{-1}(x-\mu)\right) \tag{33}$$

for some mean $\mu \in \mathbb{R}^{d_x}$ and for some covariance matrix $\Sigma \in \mathbb{R}^{d_x \times d_x}$. Possibility function (33) can be transformed from its counterpart of probability function by using the following Klir's method [22, 29],

$$\pi(x) = \frac{p(x)}{\sup_{x\prime} p(x\prime)}, \tag{34}$$

$$p(x) = \frac{\pi(x)}{\int \pi(x\prime) dx\prime}. \tag{35}$$

Given the transition possibility function $\pi(x_k|x_{k-1})$ and measurement possibility function $\pi(z_k|x_k)$, state estimation in form of possibility recursion can be presented as a similar two-

step procedure as below [25,28]:

*State prediction.*

$$\pi(x_k|z_{1:k-1}) = \sup_{x_{k-1}} \pi(x_k|x_{k-1})\pi(x_{k-1}|z_{1:k-1}) \quad (36)$$

*State update.*

$$\pi(x_k|z_{1:k}) = \frac{\pi(z_k|x_k)\pi(x_k|z_{1:k-1})}{\sup_{x_k}\pi(z_k|x_k)\pi(x_k|z_{1:k-1})} \quad (37)$$

*Remark*: State estimations in forms of sigma inference and max inference have parallel structures and differ only as follows: a) integrals are replaced by supremums (the operator of maximum is extended to supremum considering state $x_k$ is continuous) and b) probability density functions are replaced with possibility functions. It was demonstrated in [28] that the predicted and posterior mean and variance in the recursion (36) and (37) are the ones of the Kalman filter in the linear-Gaussian case [25, 28]. The underlying principle could be understood by a simple explanation as given below.

Let denote

$$\xi = x_k - F\hat{x}_{k-1|k-1} = F(x_{k-1} - \hat{x}_{k-1|k-1}) + Gw_k \quad (38)$$

then $p(x_k|x_{k-1}) = \mathcal{N}(x_k; Fx_{k-1}, GQ_kG^T)$ can be equivalently written as $p(\xi) = \mathcal{N}(\xi; 0, F\hat{\Sigma}_{k-1|k-1}F^T + GQ_kG^T)$. Then (23) can be rewritten as

$$p(x_k|z_{1:k-1}) = p(\xi)\int p(x_{k-1}|z_{1:k-1})dx_{k-1} = p(\xi) \quad (39)$$

Similarly (37) can be rewritten as

$$\pi(x_k|z_{1:k-1}) = \pi(\xi)\sup_{x_{k-1}}\pi(x_{k-1}|z_{1:k-1}) = \pi(\xi) \quad (40)$$

where $\pi(\xi) = \bar{\mathcal{N}}(\xi; 0, F\hat{\Sigma}_{k-1|k-1}F^T + GQ_kG^T)$ is equivalent to $\pi(x_k|x_{k-1}) = \bar{\mathcal{N}}(x_k; Fx_{k-1}, GQ_kG^T)$.

As we see from (39) and (40), sigma operator and max operator make functions equivalently in deriving the predicted probability/possibility for the linear-Gaussian case. In the general case, they would make functions differently.

### D. Pattern Recognition in Form of Sigma Inference

For pattern recognition in form of sigma inference, prior information such as feature-pattern mapping $p(c^i|f_k)$ and feature-measurement mapping $p(z_k|f_k)$ will be modeled by probability function. Here $c^i$ is one of $s$ known patterns from $C = \{c^1, c^2, ..., c^s\}$, $f_k$ is feature at time $k$ taking values from feature set $\mathcal{F} = \{f^1, f^2, ..., f^m\}$, $z_k$ is the measurement from an attribute sensor.

The process of pattern recognition can be formalized as the Bayesian update of posteriori probability [27,30,31]

$$p(c^i|z_{1:k}) = \frac{p(c^i|z_{1:k-1})p(z_k|c^i,z_{1:k-1})}{\sum_{c^l} p(c^l|z_{1:k-1})p(z_k|c^l,z_{1:k-1})} \quad (41)$$

where $z_{1:k} = \{z_1, z_2, ..., z_k\}$ is the feature measurement series up to time $k$, and the denominator of the right side of (41) functions as a normalization factor. Pattern likelihood $p(z_k|c^i, z_{1:k-1})$ can be propagated as follows [32]:

$$p(z_k|c^i, z_{1:k-1}) = \sum_{f_k} p(z_k|f_k, z_{1:k-1})p(f_k|c^i, z_{1:k-1}) \quad (42)$$

$$p(f_k|c^i, z_{1:k-1}) = \frac{p(f_k|z_{1:k-1})p(c^i|f_k)}{\sum_{f_l} p(f_l|z_{1:k-1})p(c^i|f_l)} \quad (43)$$

where in (42) $z_k$ and $c^i$ are supposed to be stochastically independent given $f_k$ and $z_{1:k-1}$, and in (43) $c^i$ and $z_{1:k-1}$ are assumed to be stochastically independent provided $f_k$ is known. Conditional probability $p(f_k|z_{1:k-1})$ can be learned or defined according to the involved problem.

The point estimate of pattern recognition is usually the maximum a posterior (MAP) estimate of the discrete pattern.

### E. Pattern Recognition in Form of Max Inference

Similarly, given possibility functions $\pi(c^i|f_k)$ and $\pi(z_k|f_k)$, pattern recognition in form of max inference can be formalized as below [10].

$$\pi(c^i|z_{1:k}) = \frac{\pi(c^i|z_{1:k-1})\pi(z_k|c^i,z_{1:k-1})}{\max_{c^l}\pi(c^l|z_{1:k-1})\pi(z_k|c^l,z_{1:k-1})} \quad (44)$$

$$\pi(z_k|c^i, z_{1:k-1}) = \max_{f_k} \pi(z_k|f_k, z_{1:k-1})\pi(f_k|c^i, z_{1:k-1}) \quad (45)$$

$$\pi(f_k|c^i, z_{1:k-1}) = \frac{\pi(f_k|z_{1:k-1})\pi(c^i|f_k)}{\max_{f_l} \pi(f_l|z_{1:k-1})\pi(c^i|f_l)} \quad (46)$$

where in (45) $z_k$ and $c^i$ are supposed to be possibilistically independent given $f_k$ and $z_{1:k-1}$, and in (46) $c^i$ and $z_{1:k-1}$ are supposed to be possibilistically independent given feature $f_k$. Conditional possibility $\pi(f_k|z_{1:k-1})$ can be learned or defined according to the involved problem.

*Remarks*: Compare max-inference classifier (44)~(46) with the traditional sigma-inference classifier (41)~(43), we see they use different disjunctive operations for feature $f_k$. Only the most discriminative feature is considered by the max-inference classifier whereas all features are adopted by the sigma-inference. This difference would usually lead the two classifiers to different performances [10], which is not like the linear-Gaussian case of state estimation.

## IV. RANDOM-FUZZY DUAL INTERPRETATION OF UNKNOWN QUANTITY FOR ESTIMATION & RECOGNITION

For the choice of uncertainty theory for practical applications, factors from two related viewpoints could be considered: a) interpretation of the uncertainty involved in the unknown quantity; b) functions built in with the sigma operator and the max operator, which should be the direct reference factors.

### A. The Random-Fuzzy Dual Interpretation of Unknown Quantity

The uncertainty involved in the unknown pattern, e.g., the type of non-cooperative air target being recognized, can be interpreted as either randomness or fuzziness, depending on the available prior knowledge and the perspective of cognition and modeling. Under the perspective of occurrence of objective event, the unknown pattern (target type) $X$ should take place as an objective event $x_i$ from the type set of $Y$ that consists of possible but exclusive target types. In such a case, the unknown target type should be interpreted as randomness. Under the perspective of subjective cognition of concept (event), the unknown target type $X$ can be simultaneously classified into more than one outcome from the type set of $Y$ that consists of possible but non-exclusive target types, by using the available observation data and through the process of feature extraction. In such a case, the unknown pattern should be interpreted as fuzziness. The fuzzy uncertainty arisen here has common in essence with that of the natural fuzzy concept such as Young, i.e., fuzziness is caused by the overlap of their intensions. If we regard feature as a kind of specific intermediate pattern, then we can as well interpret feature as either randomness or fuzziness.

The uncertainty involved in the unknown state, e.g., the kinematic state of non-cooperative air target being estimated, is traditionally regarded as randomness and modeled by probability. The interpretation of unknown kinematic state as randomness does make sense only if we follow the perspective of occurrence of objective event, i.e., the unknown kinematic state $X$ should take place as an objective event from the state set of $Y$. Then the possible kinematic states will be exclusive and the unknown kinematic state can take only one outcome $x_i$ from the set of $Y$. If we regard the process of state estimation as behavior of subjective cognition of the state estimator, then we can make such an interpretation that the inaccurate observation with (random) noise will cause the result of estimation to produce fuzziness because of the overlap of the intensions of different estimates.

Though the random-fuzzy dual interpretation of unknown quantity is advocated above, we take the perspective that continuous unknown quantity involved in estimation with inaccurate prior should be more appropriately regarded as randomness instead of fuzziness for three reasons below:

1) The unknown state being estimated is a continuous dynamic quantity, for which the concept of fuzziness in our understanding is not very suitable though the concept of possibility has currently been extended to cover continuous variable. Recall that the concept of fuzziness originates from discrete event (concept).

2) Estimation does not include the process of feature extraction. Therefore, it is not appropriate to interpret unknown state as fuzzy uncertainty since fuzziness stems from the overlap of the extracted features (intensions). Note that state transition as formulated by (21) is not a process of feature extraction since the involved variables $x_k$ and $x_{k+1}$ are both kinematic states but at different time steps.

3) Process noise and measurement noise, as shown in (21) and (22), should be modeled as random variables. The measurement of the target location and its error satisfy the definition of randomness, hence are suitable to be modeled as random variable. As suggested by JCGM, evaluation of uncertainty of measurement data is based on probability distributions [1]. It is as well reasonable to interpret state variable as random variable, since the measurement data is usually equal to the target location (as element of the state) plus additive noise, as can been see from (22). Otherwise, if the state variable were modeled as fuzziness, then representation of measurement noise should be transformed from probability into possibility, e.g., as in [25,28,33]. For the similar reason, process noise, i.e., the modeling error of the target state, satisfies the definition of randomness and should be modeled as random variable.

We take the perspective that discrete unknown quantity involved in recognition with insufficient (and inaccurate) prior could be better modeled as fuzziness for three reasons below:

1) The unknown pattern and feature are usually discrete static or dynamic quantity, for which the concept of fuzziness as well as randomness is applicable.

2) Recognition usually includes the process of feature extraction. This process consists of progressive steps of transformation between different discrete features, which cause the overlap of the extracted intensions hence the fuzziness.

3) Measurement noise, as shown in (22), should be modeled as random variable. Nevertheless, the extracted feature and the unknown pattern should be modeled as fuzzy variable.

Overall, the recognition of discrete pattern could be better regarded as a cognition process, where the involved unknown uncertainty should be interpreted as fuzzy uncertainty and the max inference could be applied. Whereas the estimation of continuous unknown state could be better regarded as a problem of stochastic filtering, where the involved unknown uncertainty should be interpreted as random uncertainty and the sigma inference could be used.

*B. On the Choice of Sigma-Max Inference*

We now discuss the choice of sigma and max operators by considering their build-in functions, typically as encoded by (6) and (7), respectively. The difference of them lies in that during the process of uncertainty inference, sigma operator uses all possible values of the unknown intermediate variable of $Y$, whereas max operator selects only one value that would make the combined possibility $\pi(z_k|x_i)$ having the maximum of one. This property of possibility originates from (2) which claims that at least one of the elements of $\Psi$ should be fully possible. On the other side, it would be our understanding that during the processing of subjective inference of human brain, the sigma normalization of probability is generally not a requirement but the max normalization like that of possibility would usually be required by default. That is, during the process of subjective inference such as encoded by (6) we usually would accept such a default assumption that there always exists a route with maximum possibility of one among all possible routes consisting of node variables $X$, $Y$ and Z.

As discussed before, estimation and recognition are both like the problem of subjective cognition, where the involved unknown (intermediate) quantity at one instant would in fact take only one value. Taking the most likely value as the approximation of the unknown (intermediate) quantity is therefore the natural and best option, and can be satisfied by the max operator. From this point of view, max operator will be the right choice for the problem of pattern recognition where discrete feature is usually involved. Besides, it usually makes no sense to approximate the effectiveness of a possible value of (intermediate) discrete feature by the composite effectiveness of multiple values, such as the case of (7). Nevertheless, it does make sense to approximate the effectiveness of a possible value of a continuous state in the linear-Gaussian case by the composite effectiveness of multiple values, as can be seen from (39). Therefore, sigma operator should be the right choice for the problem of estimation of continuous state considering that the process/measurement noises are suggested to be interpreted as random uncertainty.

For hybrid estimation of jump Markov System, where the involved unknown continuous state should be modeled as randomness whereas the unknown discrete mode should be modeled as fuzziness, the sigma-max hybrid uncertainty inference should be applied, which will be discussed in detail as in section 5.

## V. A Demonstration Example of IMM Filter

Over the last three decades, the IMM filter has been gradually recognized as an efficient estimation method for linear systems with Markov switching coefficients [34-38]. The IMM filter belongs to a family of estimation methods under the name of multiple-model state estimation or hybrid state estimation, which also include the generalized pseudo-Bayesian (GPB) algorithm [39,40] and the Viterbi algorithm (VA) [41]. Being multiple-model estimator, the dynamics of the jump Markov system are represented through a finite set of models governed by a Markov chain, with each model describing continuous state evolution and the Markov chain describing discrete mode transition. Multiple-model estimator shows up in the form of a bank of interacting Kalman filters, with each filter matched to a mode of the jump Markov system. This work will mix the use of mode and model considering that sometimes one of them is more accurate and sometimes either of them makes sense.

To be the optimal Bayesian multiple-model estimator, all historic model-sequences of the Markov chain should be taken into consideration from the initial stage. As the number of such "histories or hypotheses" grows exponentially with time, the optimal solution is intractable [20,41]. By selecting a single "most probable history", the Viterbi algorithm was developed in [41]. In parallel, solutions based on "hypotheses merging" seem to exhibit more advantages. For the GPB $(d + 1)$ algorithm, model hypotheses are merged, immediately after measurement update, to one moment-matched Gaussian hypothesis with a fixed depth of $d$ [20,41]. For the IMM filter, the timing of hypotheses merging, with a depth of $d = 1$, is moved to before filtering. The IMM filter performs almost as well as the GPB2 algorithm, while its computational load is about that of the GPB1 algorithm [20]. The merits of excellent performance and linear model-complexity bring the IMM filter a reputation of over 2000 citations of the work [20] so far.

The above multiple-model estimators are all based on probability theory and the derivative subjects of statistics and stochastic process, where uncertainties related to continuous states and discrete modes are all modeled as random uncertainty. This traditional handling of the uncertainties under a pure probability perspective, perhaps, needs to have a re-examination. Let us start our re-examination with the term of "optimal". As mentioned above that the optimal Bayesian multiple-model estimator needs to consider all historic model-sequences of the Markov chain, which grows exponentially over time. However, the jump Markov system of the real world should at a certain moment (sample time) be in an un-known but fixed mode, instead of multiple possible modes. To this end, probability is not very suitable for modeling of the jump Markov system. And the "optimal" of the multiple-model estimator is only in a sense of probability, but not relative to the real world. For the jump Markov system of the real world, we in one hand need to use a bank of interacting Kalman filters to cover all its possible modes; on the other hand, we need to decide at each moment a single mode that is most possible.

To overcome the weakness mentioned above, the HIMM filter is to be developed below, for which continuous state evolution of the jump Markov system is as usual modeled as model-conditioned stochastic system but the uncertainty related to system mode is regarded as fuzziness and discrete mode transition is modeled by a possibility transition matrix instead of a probability matrix. The HIMM filter was developed by using the sigma-max hybrid uncertainty inference, and is more like the IMM filter than like the Viterbi algorithm. Compared with the classic IMM filter, simulation results in section 6 show that the HIMM filter has significantly better performance due to its peculiarity of hard decision of system model and a faster response to the transition of discrete mode.

### A. Problem Formulation

Consider the following jump Markov system [35]

$$x_k = F(r_k)x_{k-1} + G(r_k)w_k \quad (47)$$

$$z_k = H(r_k)x_k + v_k \quad (48)$$

where system mode index $r_k$ is described by a finite Markov chain, which takes values from model set $\mathcal{M} = \{1,2 \dots, M\}$. $F(r_k)$, $G(r_k)$ and $H(r_k)$ are known matrices for each mode index $r_k$.

For the classic IMM filter, the transition of system mode is described by the transition probability matrix

$$\mathcal{P} = [p_{ij}]_{M \times M} = [p(r_k = j | r_{k-1} = i)]_{M \times M} \quad (49)$$

where $p_{ij}$ denotes the transition probability from mode $i$ to mode $j$ and it satisfies $\sum_{j=1}^{M} p_{ij} = 1$ for any $i \in \mathcal{M}$. Jump Markov system formulated by (47)~(49) is a stochastic linear system.

According to the perspective presented in this paper, the system mode $r_k$, which is an unknown discrete process, could be better interpreted as fuzziness and modeled by the transition possibility matrix

$$\Pi = [\pi_{ij}]_{M \times M} = [\pi(r_k = j | r_{k-1} = i)]_{M \times M} \quad (50)$$

where $\pi_{ij}$ denotes the transition possibility from mode $i$ to mode $j$ and it satisfies $\max_{j \in \mathcal{M}} \pi_{ij} = 1$ for any $i \in \mathcal{M}$. We name jump Markov system formulated by (47), (48) and (50) as fuzzy jump Markov system, and that of (47)~(49) as random jump Markov system.

The problem considered is formulated as follows: Given the hybrid uncertainty linear system (47), (48) and (50), derive a multiple-model estimator to compute the posterior distribution $p(x_k|z_{1:k})$ and/or its first and second moments $\hat{x}_{k|k}$ and $\hat{\Sigma}_{k|k}$, by using the sigma-max inference.

*B. The Hybrid IMM filter*

The HIMM filter consists of the following four steps, the derivation of which is left to the Appendix.

**Step 1**: Model Interaction

Starting with the mode possibility $\pi(r_{k-1} = l|z_{1:k-1})$, the mode possibility $\pi(r_k = j|z_{1:k-1})$ after interaction and then the move-in mode possibility $\pi(r_{k-1} = l|r_k = j, z_{1:k-1})$ can be figured out by (51) and (52) below.

$$\pi(r_k = j|z_{1:k-1}) = \max_{l \in \mathcal{M}} \pi(r_k = j|r_{k-1} = l)\pi(r_{k-1} = l|z_{1:k-1}) \quad (51)$$

$$\pi(r_{k-1} = l|r_k = j, z_{1:k-1}) = \frac{\pi(r_k = j|r_{k-1} = l)\pi(r_{k-1} = l|z_{1:k-1})}{\pi(r_k = j|z_{1:k-1})} \quad (52)$$

Note that for the IMM, (51) and (52) will be replaced by [20,26]

$$p(r_k = j|z_{1:k-1}) = \sum_{l=1}^{M} p(r_k = j|r_{k-1} = l)p(r_{k-1} = l|z_{1:k-1}) \quad (53)$$

$$p(r_{k-1} = l|r_k = j, z_{1:k-1}) = \frac{p(r_k = j|r_{k-1} = l)p(r_{k-1} = l|z_{1:k-1})}{p(r_k = j|z_{1:k-1})} \quad (54)$$

With the move-in possibility $\pi(r_{k-1} = l|r_k = j, z_{1:k-1})$, the mean $\hat{x}^j_{k-1|k-1}$ and the associated covariance $\hat{\Sigma}^j_{k-1|k-1}$, we can compute the mixed initial condition for the filter by (55)~(57) below. Eq. (55) searches among model set $\mathcal{M}$ for the most possible mode $l$. Eq. (56) assigns every mixed mean $\hat{x}^{0j}_{k-1|k-1}$ with the same mean $\hat{x}^l_{k-1|k-1}$ that has the maximum mode possibility.

$$l = \arg\max_{l \in \mathcal{M}} \pi(r_{k-1} = l|r_k = j, z_{1:k-1}) \quad (55)$$

$$\hat{x}^{0j}_{k-1|k-1} = \hat{x}^l_{k-1|k-1} \quad (56)$$

$$\hat{\Sigma}^{0j}_{k-1|k-1} = \hat{\Sigma}^j_{k-1|k-1} + [\hat{x}^{0j}_{k-1|k-1} - \hat{x}^l_{k-1|k-1}][..]^T \quad (57)$$

For the IMM, (55)~(57) will be replaced by [20,26]

$$\hat{x}^{0j}_{k-1|k-1} = \sum_{l=1}^{M} p(r_{k-1} = l|r_k = j, z_{1:k-1}) \hat{x}^l_{k-1|k-1} \quad (58)$$

$$\hat{\Sigma}^{0j}_{k-1|k-1} = \sum_{l=1}^{M} p(r_{k-1} = l|r_k = j, z_{1:k-1}) \{\hat{\Sigma}^j_{k-1|k-1} + [\hat{x}^{0j}_{k-1|k-1} - \hat{x}^l_{k-1|k-1}][..]^T\} \quad (59)$$

**Step 2**: Model-conditioned Filtering

Each of the $M$ pairs $\hat{x}^{0j}_{k-1|k-1}$ and $\hat{\Sigma}^{0j}_{k-1|k-1}$ is used as input to the Kalman filter matched to model $j$. As indicated by (60) and (61), time extrapolation yields $\hat{x}^j_{k|k-1}$ and $\hat{\Sigma}^j_{k|k-1}$, and then measurement update gives $\hat{x}^j_{k|k}$ and $\hat{\Sigma}^j_{k|k}$.

$$\hat{x}^{0j}_{k-1|k-1} \Rightarrow \hat{x}^j_{k|k-1} \Rightarrow \hat{x}^j_{k|k} \quad (60)$$

$$\hat{\Sigma}^{0j}_{k-1|k-1} \Rightarrow \hat{\Sigma}^j_{k|k-1} \Rightarrow \hat{\Sigma}^j_{k|k} \quad (61)$$

For the IMM, model-conditioned filtering will be the same as the HIMM.

**Step 3**: Model Possibility Update

The updated model possibility becomes

$$\pi(r_k = j|z_{1:k}) = \frac{\Lambda^j_k \pi(r_k = j|z_{1:k-1})}{\max_{l \in \mathcal{M}} \Lambda^l_k \pi(r_k = l|z_{1:k-1})} \quad (62)$$

where $\Lambda^j_k = p(z_k|r_k = j, z_{1:k-1})$ is the likelihood function of model $j$ given observation $z_k$.

For the IMM, (62) will be replaced by [20,26]

$$p(r_k = j|z_{1:k}) = \frac{\Lambda^j_k p(r_k = j|z_{1:k-1})}{\sum_{l=1}^{M} \Lambda^l_k p(r_k = l|z_{1:k-1})} \quad (63)$$

**Step 4**: Estimation Output

The final output from the HIMM filter is given by

$$j = \arg\max_{j \in \mathcal{M}} \pi(r_k = j|z_{1:k}) \quad (64)$$

$$\hat{x}_{k|k} = \hat{x}^j_{k|k} \quad (65)$$

$$\hat{\Sigma}_{k|k} = \hat{\Sigma}^j_{k|k} \quad (66)$$

For the IMM, (64)~(66) will be replaced by [20,26]

$$\hat{x}_{k|k} = \sum_{l=1}^{M} p(r_k = l|z_{1:k}) \hat{x}^l_{k|k} \quad (67)$$

$$\hat{\Sigma}_{k|k} = \sum_{l=1}^{M} p(r_k = l|z_{1:k}) \hat{\Sigma}^l_{k|k} \quad (68)$$

*Remark*: The HIMM filter has a structure that is parallel to the classic IMM filter as presented in [20]. During a circle of the algorithm, the HIMM runs in parallel $M$ model-matched filters and only one filter with the maximum model possibility is fully in charge of the filtering. Whereas for the IMM, each model-matched filter will share a certain probability of responsibility of the filtering work. Be aware that the HIMM is not equivalent to a variation of the IMM filter that would output results of the model-conditioned filter with the largest model probability as the final global estimation.

## VI. SIMULATIONS

Two scenarios are provided to compare the HIMM and the IMM for tracking a maneuvering target, where simulated target measurements are from a short-range fire control radar and a long-range surveillance radar, respectively.

### A. Data Generation

Scenario 1: Short-range fire control radar tracking with sampling interval of measurement update $T = 0.2s$.

A target flies in 3-dimension space with initial state vector $\boldsymbol{x_0} = [x_0, \dot{x}_0, \ddot{x}_0, y_0, \dot{y}_0, \ddot{y}_0, z_0, \dot{z}_0, \ddot{z}_0]^T$ given by

$$[x_0, y_0, z_0] = [12km,\ 8km,\ 1km]$$
$$[\dot{x}_0, \dot{y}_0, \dot{z}_0] = [-100m/s, -100m/s, 0m/s]$$
$$[\ddot{x}_0, \ddot{y}_0, \ddot{z}_0] = [0m/s^2, 0m/s^2, 0m/s^2]$$

The target accelerates from the 81$^{st}$ sample with $[\ddot{x}_0, \ddot{y}_0, \ddot{z}_0] = [-30m/s^2, -50m/s^2, 0m/s^2]$ until the 130$^{th}$ sample. It maintains the velocity at the 131$^{st}$ sample to the 200$^{th}$ sample. The fire control radar measures target range and bearings with alternative accuracies as listed in Table 1, where $\sigma_\alpha$, $\sigma_\beta$ and $\sigma_\gamma$ indicate the standard deviations of azimuth, elevation and range, respectively.

TABLE I
RADAR MEASUREMENT ACCURACY

|  | $\sigma_\alpha$ | $\sigma_\beta$ | $\sigma_\gamma$ |
|---|---|---|---|
| Fire control radar | 0.1° (0.2°) | 0.1° (0.2°) | 10m (20m) |
| Surveillance radar | 0.9° (1.8°) | 0.9° (1.8°) | 100m (200m) |

Scenario 2: Long-range surveillance radar tracking with sampling interval of measurement update $T = 2s$.

The initial state vector of the target is given by

$$[x_0, y_0, z_0] = [120km,\ 80km,\ 20km]$$
$$[\dot{x}_0, \dot{y}_0, \dot{z}_0] = [-100m/s, -100m/s, 0m/s]$$
$$[\ddot{x}_0, \ddot{y}_0, \ddot{z}_0] = [0m/s^2, 0m/s^2, 0m/s^2]$$

The target accelerates from the 31$^{st}$ sample with $[\ddot{x}_0, \ddot{y}_0, \ddot{z}_0] = [-30m/s^2, -50m/s^2, 0m/s^2]$ until the 40$^{th}$ sample. It maintains the velocity at the 41$^{st}$ sample to the 80$^{th}$ sample. The accuracies of the surveillance radar are given in Table 1.

For both scenarios, process noise $w_k$ is applied to the generated target trajectory, which is white zero-mean Gaussian with deviation $\sigma_{w_k} = 3m/s^2$.

### B. Tracker Parameters

Before being processed with a tracker, measurements from the radar are first converted to Cartesian coordinates using the standard conversion method [2,42]. For unbiased conversion methods, the readers may refer to [43, 44]. For both the HIMM and the IMM, two dynamic models are selected, which are the discrete white noise acceleration (DWNA) model and the discrete Wiener process acceleration (DWPA) model [2,26].

Denote $F = \mathrm{diag}(F_s, F_s, F_s)$, $G = \mathrm{diag}(G_s, G_s, G_s)$ and $H = \mathrm{diag}(H_s, H_s, H_s)$, then for the DWNA model

$$F_s = \begin{pmatrix} 1 & T \\ 0 & 1 \end{pmatrix}, \quad G_s = \begin{pmatrix} \frac{1}{2}T^2 \\ T \end{pmatrix}, \quad H_s = (1\ 0).$$

For the DWPA model

$$F_s = \begin{pmatrix} 1 & T & \frac{1}{2}T^2 \\ 0 & 1 & T \\ 0 & 0 & 1 \end{pmatrix}, \quad G_s = \begin{pmatrix} \frac{1}{2}T^2 \\ T \\ 1 \end{pmatrix}, \quad H_s = (1\ 0\ 0).$$

The two models have equal initial probabilities/possibilities with transition matrices fine-tuned and given by

$$\mathcal{P} = [p_{ij}]_{M \times M} = \begin{pmatrix} 0.95 & 0.05 \\ 0.05 & 0.95 \end{pmatrix},$$

$$\Pi = [\pi_{ij}]_{M \times M} = \begin{pmatrix} 1 & 1/2 \\ 1/2 & 1 \end{pmatrix}.$$

The state estimation of the model-matched Kalman filters are initialized with two or three points difference [2].

### C. Design of Experiment Groups

For both scenarios, four groups of simulation parameters as below are designed. Note that parameters in the bracket are for surveillance radar. Group 1 represents the case that model parameters exactly match data parameters (parameters for data generation). Groups 2~3 indicate that model parameters are conservative relative to data parameters. Group 4 indicates that model parameters are optimistic relative to data parameters.

**Group 1:**
Data parameters:
$\sigma_{w_k} = 3m/s^2$
$\sigma_\alpha = \sigma_\beta = 0.1°\ (0.9°)$, $\sigma_\gamma = 10m\ (100m)$
Model parameters:
$\sigma_{w_k} = 3m/s^2$
$\sigma_\alpha = \sigma_\beta = 0.1°\ (0.9°)$, $\sigma_\gamma = 10m\ (100m)$

**Group 2:**
Data parameters: the same as Group 1
Model parameters:
$\sigma_{w_k} = 3m/s^2$
$\sigma_\alpha = \sigma_\beta = 0.15°\ (1.35°)$, $\sigma_\gamma = 15m\ (150m)$

**Group 3:**
Data parameters: the same as Group 1

Model parameters:
$\sigma_{w_k} = 3\text{m/s}^2$
$\sigma_\alpha = \sigma_\beta = 0.2°\ (1.8°),\ \sigma_\gamma = 20m\ (200m)$

**Group 4:**
Data parameters:
$\sigma_{w_k} = 3\text{m/s}^2$
$\sigma_\alpha = \sigma_\beta = 0.2°\ (1.8°),\ \sigma_\gamma = 20m\ (200m)$
Model parameters:
$\sigma_{w_k} = 1\text{m/s}^2$
$\sigma_\alpha = \sigma_\beta = 0.1°\ (0.9°),\ \sigma_\gamma = 10m\ (100m)$

*D. Simulation Results*

Simulation results are based on the average of 100 Monte Carlo runs. The root mean-square error (RMSE) [21] as defined below is used to measure the position estimation error in axis of *x*, *y* or *z*.

$$\text{RMSE} = \sqrt{\frac{1}{M_c}\sum_{s=1}^{M_c}(x^s - \hat{x}^s)^2}$$

where $x^s$ and $\hat{x}^s$ are the true and estimated positions for the $s^{\text{th}}$ Monte Carlo run, respectively.

Results of Group 1 are shown in Figs. 2~11, whereas results of Groups 2~4 are to be presented only by text descriptions. Note that measurement error is also plotted in those figures to provide as reference.

Group 1: data/model parameters are exactly matched. For Scenario 1 of fire control radar tracking, Figs. 2~6 show that the HIMM has significantly better performance in all axes, whenever the target is maneuvering or not, than the classic IMM. The estimation errors of the *z* axis arise at the 81$^{\text{st}}$ sample, which should be caused by coupling across different axes of *x*, *y* and *z* since there is no acceleration along the *z* axis. Figs. 5 and 6 show that the HIMM has a larger model noise, but with a faster response to the mode switch. The average cross-times of model switch of the HIMM and IMM are the 83.77$^{\text{th}}$ scan and the 85.92$^{\text{th}}$ scan, respectively.

For Scenario 2 of surveillance radar tracking, Figs. 7~11 show that the two methods have competitive yet complementary performance. The HIMM is better before and by the end of the maneuvering. During the maneuvering the IMM is slightly better. Figs. 10 and 11 show that the HIMM responds slightly faster to the mode switch (31.88: 32.28). The probable reason that the IMM has a slight better accuracy for maneuvering period is that a better estimation of the HIMM at scan $(k-1)$ will lead to a bigger prediction error at the scan $k$ when mode-switch happened. The weak advantage of faster model-switch of the HIMM would have been neutralized by this disadvantage. We in the experiments increased the model parameter of $\sigma_{w_k}$ for the HIMM from $3\text{m/s}^2$ to $5\text{m/s}^2$, which would slightly decrease the accuracy of the HIMM for the non-maneuvering period, and we observed that the HIMM outperforms the IMM.

Group 2: the measurement noise of the model parameters is increased by 50% compared to that of Group 1, with all other parameters the same as those of Group 1. For Scenario 1, the HIMM is significantly better before and by the end of the maneuvering. During the maneuvering, the HIMM and the IMM are competitive. For Scenario 2, before the maneuvering, the HIMM is better. During the maneuvering, the IMM is better. By the end of the maneuvering, they are competitive. For both scenarios, the two algorithms run stably with good filtering effects and model switches observed.

Group 3: the measurement noise of the model parameters is increased by 100% compared to that of Group 1, with all other parameters the same as those of Group 1. For both scenarios, the two algorithms run stably with good filtering effects and model switches observed. The HIMM is better before the maneuvering and significantly worse during the maneuvering. By the end of the maneuvering, the HIMM is better than (Scenario 1) or competitive with (Scenario 2) the IMM. During the maneuvering, models of the HIMM switch frequently, which means with remarkably mismatched parameters, the HIMM cannot make correct model decision.

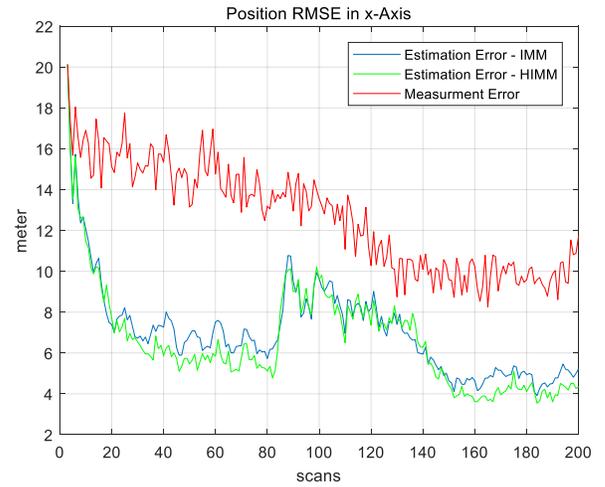
Fig. 2. Scenario 1: Position error on the *x* axis.

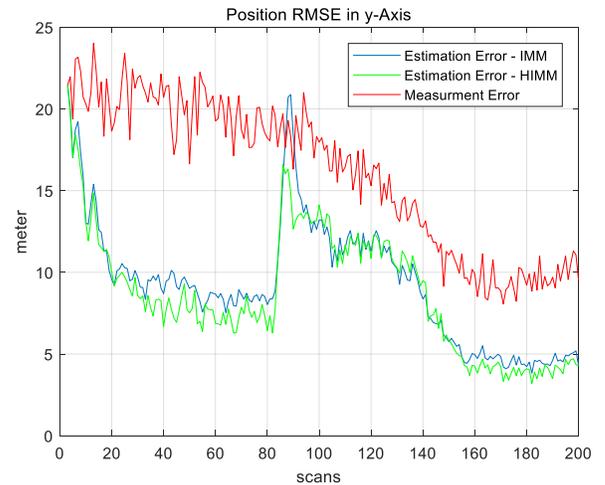
Fig. 3. Scenario 1: Position error on the *y* axis.

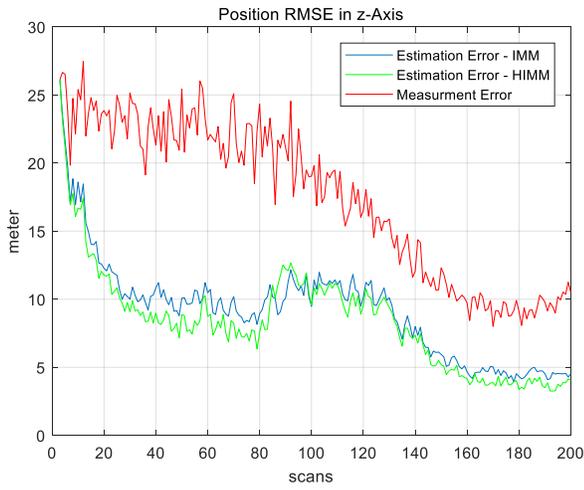
Fig.4. Scenario 1: Position error on the *z* axis.

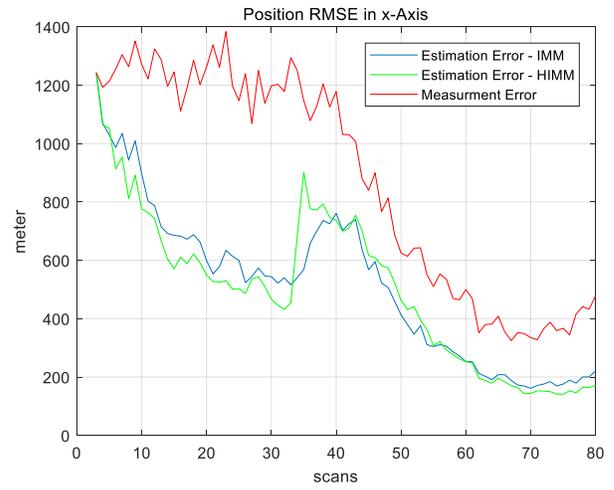
Fig. 7. Scenario 2: Position error on the *x* axis.

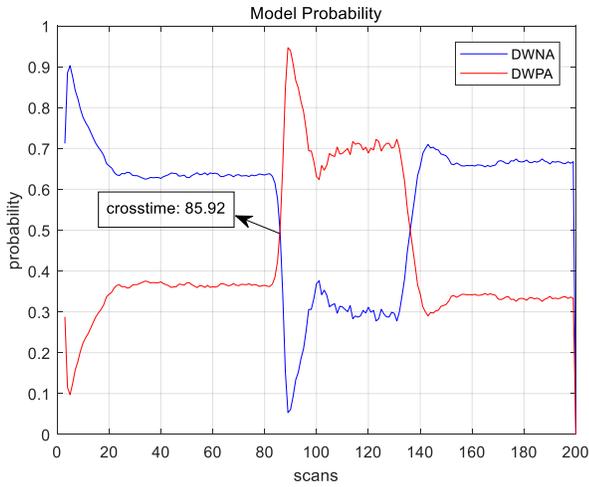
Fig. 5. Scenario 1: Model probabilities of the IMM.

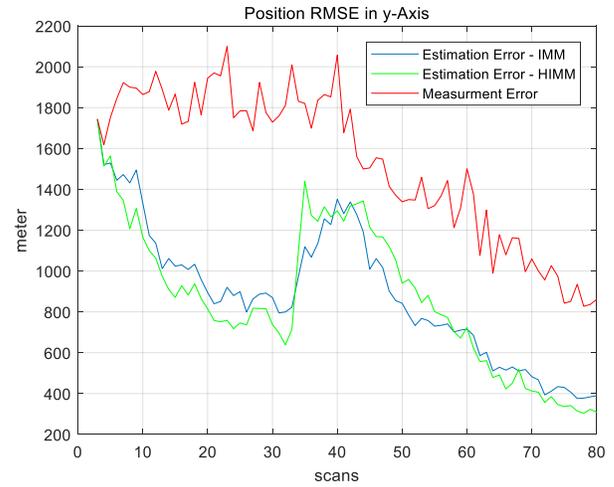
Fig. 8. Scenario 2: Position error on the *y* axis.

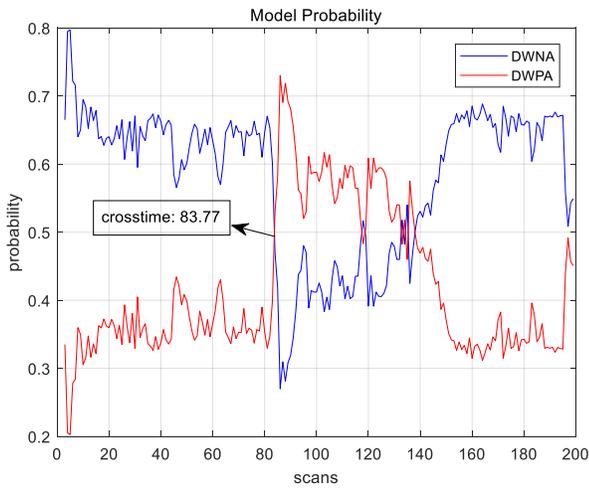
Fig. 6. Scenario 1: Model probabilities of the HIMM.

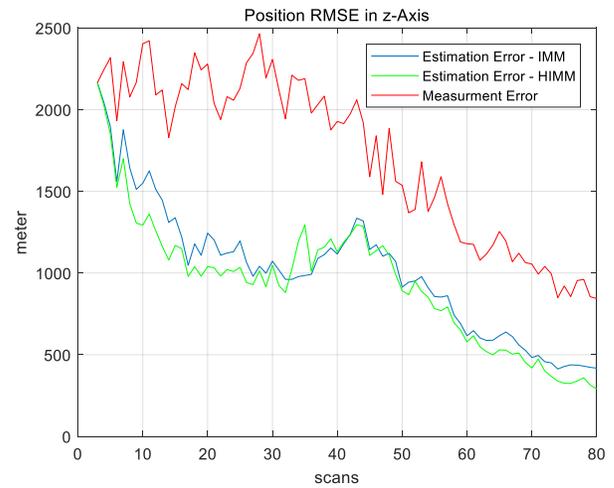
Fig. 9. Scenario 2: Position error on the *z* axis.

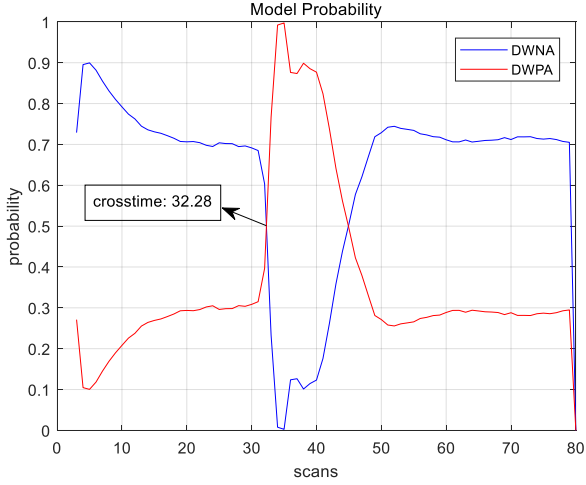

Fig. 10. Scenario 2: Model probabilities of the IMM.

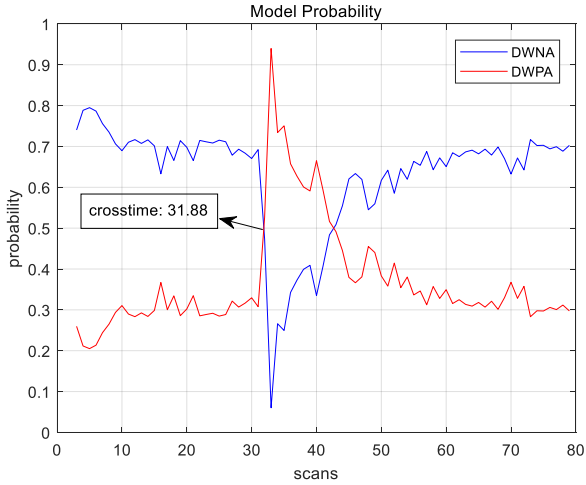

Fig. 11. Scenario 2: Model probabilities of the HIMM.

Group 4: model parameters are set to be remarkably optimistic compared to data parameters. For Scenario 1, the two algorithms generally run stably, but occasionally suffer numerical problem in computing the model likelihood. Model switches disappeared and the IMM always identifies the system mode as DWNA model whereas the HIMM always identifies the mode as DWPA model. The overall performances of the two methods are competitive. Once the maneuvering starts, the estimation error of the IMM filter would approach the measurement error, which means the IMM has lost the function of filtering. For Scenario 2, the two algorithms run stably with competitive performance, but model switches disappeared as well. The IMM/HIMM always identifies the mode as DWNA/DWPA model, respectively, as well. Before the maneuvering, the IMM is better. During the maneuvering, the HIMM is better. By the end of maneuvering, they are competitive.

*E. Discussions*

By experiments of Group 1 when data/model parameters are exactly matched, we see the HIMM filter has significantly better performance than the classic IMM filter. A better performance for the period of non-maneuvering can be attributed to the hard decision of system model of the HIMM, and a better performance of maneuvering tracking may be due to the faster response of the HIMM to the mode switch.

By experiments of Groups 2~4 when data/model parameters suffer from different forms of mismatch, we see both methods exhibit robust performance and good filtering effects. The two methods are generally competitive, with their respective merits and shortages for different simulation conditions.

Our experiments also show that a variation of the IMM, which outputs results of the model-conditioned filter with the largest model probability as the final global estimation, has a very close performance with the IMM.

We did not compare the HIMM with other improvements or extensions, e.g., as presented in [21,35,45-47], of the classic IMM, which are all formulated in the framework of probability theory. We consider the HIMM as a counterpart of the IMM and most of those improvements or extensions could be applied to the HIMM, as well.

## VII. CONCLUSION

The updated version of the well-known IMM filter demonstrated our perspective that continuous unknown quantity involved in estimation should be more appropriately modeled as randomness and handled by sigma inference, whereas discrete unknown quantity involved in recognition could be better modeled as fuzziness and handled by max inference. For hybrid estimation of jump Markov System, where both continuous state and discrete mode are involved, the sigma-max hybrid inference should be applied. Our perspective presented in this work can also find support from the example in [10] of target recognition using simulated data from an electronic support measure (ESM), which shows the sigma-max classifier with heterogeneous information fusion performs significantly better than the traditional sigma-inference classifier. It remains our ongoing efforts to find more applications of the sigma-max inference in other areas including the booming area of machine learning [48,49], along with the purpose of supporting or refining our perspective of random-fuzzy dual interpretation of unknown quantity.

APPENDIX. DERIVATION OF THE HIMM FILTER

**Step 1**: Model Interaction

Following (17), the mixed state distribution can be figured out by

$$p(x_{k-1}|r_k = j, z_{1:k-1}) = \frac{1}{\beta} p^+(x_{k-1}|r_k = j, z_{1:k-1})$$
$$= \frac{1}{\beta} \max_{l \in \mathcal{M}} p\pi(x_{k-1}, r_{k-1} = l|r_k = j, z_{1:k-1})$$
$$= \frac{1}{\beta} \max_{l \in \mathcal{M}} p(x_{k-1}|r_{k-1} = l, z_{1:k-1})\pi(r_{k-1} = l|r_k = j, z_{1:k-1})$$
(A-1)

Let

$$l = \arg \{\max_{l \in \mathcal{M}} p(x_{k-1}|r_{k-1} = l, z_{1:k-1}) \\ \times \pi(r_{k-1} = l|r_k = j, z_{1:k-1})\}, \quad \text{(A-2)}$$

then we have

$$p(x_{k-1}|r_k = j, z_{1:k-1}) = \frac{1}{\beta} p(x_{k-1}|r_{k-1} = l, z_{1:k-1})\pi(r_{k-1} = l|r_k = j, z_{1:k-1}). \quad \text{(A-3)}$$

As we can see from (A-3), in our model world based on the sigma-max inference, the computation of the mixed distribution $p(x_{k-1}|r_k = j, z_{1:k-1})$ would possibly contain an information fusion between continuous state $x_{k-1}$ and discrete mode $r_{k-1}$. To make the HIMM be implementable in form of the Kalman filter, we suppose (A-2) can be approximated by

$$l = \arg\max_{l \in \mathcal{M}} \pi(r_{k-1} = l|r_k = j, z_{1:k-1}), \quad \text{(A-4)}$$

then $\pi(r_{k-1} = l|r_k = j, z_{1:k-1}) = 1$ and $\beta = 1$, and we have

$$p(x_{k-1}|r_k = j, z_{1:k-1}) = p(x_{k-1}|r_{k-1} = l, z_{1:k-1}). \quad \text{(A-5)}$$

The mixed mode possibility is computed by

$$\pi(r_k = j|z_{1:k-1}) = \max_{l \in \mathcal{M}} \pi(r_k = j, r_{k-1} = l|z_{1:k-1})$$
$$= \max_{l \in \mathcal{M}} \pi(r_k = j|r_{k-1} = l)\pi(r_{k-1} = l|z_{1:k-1}) \quad \text{(A-6)}$$

The move-in mode possibility is calculated by

$$\pi(r_{k-1} = l|r_k = j, z_{1:k-1}) = \frac{\pi(r_k = j|r_{k-1} = l)\pi(r_{k-1} = l|z_{1:k-1})}{\pi(r_k = j|z_{1:k-1})} \quad \text{(A-7)}$$

**Step 2**: Model-conditioned Filtering

The model-conditioned posterior distribution is updated as

$$p(x_k|r_k = j, z_{1:k})$$
$$= \frac{p(z_k|x_k, r_k = j, z_{1:k-1})p(x_k|r_k = j, z_{1:k-1})}{p(z_k|r_k = j, z_{1:k-1})}$$
$$= \frac{p(z_k|x_k)p(x_k|r_k = j, z_{1:k-1})}{\int p(z_k|x_k)p(x_k|r_k = j, z_{1:k-1})dx_k}, \quad \text{(A-8)}$$

where the predicted state is computed by

$$p(x_k|r_k = j, z_{1:k-1}) = \int p(x_k|x_{k-1}, r_k = j)p(x_{k-1}|r_k = j, z_{1:k-1})dx_{k-1} \quad \text{(A-9)}$$

**Step 3**: Model Possibility Update

By the possibility update in form of (19), the updated model possibility becomes

$$\pi(r_k = j|z_{1:k}) = \frac{p(z_k|r_k = j, z_{1:k-1})\pi(r_k = j|z_{1:k-1})}{\max_{l \in \mathcal{M}} p(z_k|r_k = l, z_{1:k-1})\pi(r_k = l|z_{1:k-1})}$$
$$= \frac{\Lambda_k^j \pi(r_k = j|z_{1:k-1})}{\max_{l \in \mathcal{M}} \Lambda_k^l \pi(r_k = l|z_{1:k-1})} \quad \text{(A-10)}$$

where $\Lambda_k^j = p(z_k|r_k = j, z_{1:k-1})$ is the likelihood function of model $j$ given observation $z_k$.

**Step 4**: Estimation Output

Following (17), the posterior distribution $p(x_k|z_{1:k})$ can be expanded as

$$p(x_k|z_{1:k}) = \frac{1}{\beta} p^+(x_k|z_{1:k})$$
$$= \frac{1}{\beta} \max_{j \in \mathcal{M}} p\pi(x_k, r_k = j|z_{1:k})$$
$$= \frac{1}{\beta} \max_{j \in \mathcal{M}} p(x_k|r_k = j, z_{1:k})\pi(r_k = j|z_{1:k}). \quad \text{(A-11)}$$

Let

$$j = \arg\max_{j \in \mathcal{M}} p(x_k|r_k = j, z_{1:k})\pi(r_k = j|z_{1:k}), \quad \text{(A-12)}$$

then we have

$$p(x_k|z_{1:k}) = \frac{1}{\beta} p(x_k|r_k = j, z_{1:k})\pi(r_k = j|z_{1:k}). \quad \text{(A-13)}$$

Similarly, we suppose (A-12) can be simplified as

$$j = \arg\max_{l \in \mathcal{M}} \pi(r_k = j|z_{1:k}), \quad \text{(A-14)}$$

then we have

$$p(x_k|z_{1:k}) = p(x_k|r_k = j, z_{1:k}). \quad \text{(A-15)}$$

With distributions in (A-5), (A-8), (A-9) and (A-15) replaced by their means and covariances, respectively, we derived the HIMM filter. The mean $\hat{x}_{k|k}$ and the covariance $\hat{\Sigma}_{k|k}$ of $p(x_k|z_{1:k})$ in (A-15), for example, can be computed by

$$\hat{x}_{k|k} = \int x_k p(x_k|r_k = j, z_{1:k})dx_k = \hat{x}_{k|k}^j \quad \text{(A-16)}$$

$$\hat{\Sigma}_{k|k} = E\left[(x_k - \hat{x}_{k|k}^j)(..)^T \Big| z_{1:k}\right] = \hat{\Sigma}_{k|k}^j \quad \text{(A-17)}$$

### References

[1] Evaluation of Measurement Data—Guide to the Expression of Uncertainty in Measurement, (GUM 1995 With Minor Corrections), document JCGM 100:2008, Joint Committee for Guides in Metrology, 2008.
[2] Y. Bar-Shalom, X. R. Li and T. Kirubarajan, *Estimation with Applications to Tracking and Navigation: Theory, Algorithms and Software*, Wiley-Interscience, 2001.
[3] A. K. Jain, R. P. W. Duin and J. Mao, Statistical pattern recognition: a review, *IEEE Transactions on Pattern Analysis and Machine Intelligence*, 22(1), Jan. 2000: 4-37, doi: 10.1109/34.824819.
[4] W. Mei, M. Li, Y.Z. Cheng and L.M. Liu, Sigma-max system induced from randomness and fuzziness, arXiv preprint arXiv:2110.07722 [cs.AI] (2021), https://arxiv.org/abs/2110.07722.
[5] https://www.britannica.com/topic/intension (2019), Accessed 25 August 2019.
[6] S. H. Unger, Pattern Detection and Recognition, *Proceedings of the IRE*, 47(10), Oct. 1959: 1737-1752.
[7] P. Eykhoff, Process parameter and state estimation, Automatica, 4(4) 1968: 205-233.
[8] W. Mei, Probability/Possibility Systems for Modeling of Random/Fuzzy Information with Parallelization Consideration, International Journal of Fuzzy Systems, April 2019, DOI: 10.1007/s40815-019-00627-9
[9] W. Mei, Formalization of Fuzzy Control in Possibility Theory via Rule Extraction, IEEE Access, July 2019, DOI: 10.1109/ACCESS.2019.2928137

[10] W. Mei, L.M. Liu, J. Dong, The Integrated Sigma-Max System and Its Application in Target Recognition, Information Sciences. 555, May 2021: 198-214, DOI: 10.1016/j.ins.2020.12.054
[11] L. A. Zadeh, Fuzzy sets, Information and control, 8 (1965): 338–353.
[12] H.-J. Zimmermann, Fuzzy Set Theory—And Its Applications, 4th ed. New York, NY, USA: Springer, 2001.
[13] Y. Jin, W. Cao, M. Wu, et al, Simplified outlier detection for improving the robustness of a fuzzy model, SCIENCE CHINA Information Sciences 63, 149201 (2020)
[14] K. Tanaka and H. O. Wang, Fuzzy Control Systems Design and Analysis: A Linear Matrix Inequality Approach. Hoboken, NJ, USA: Wiley, 2001.
[15] W. Sun, S.-F. Su, J. Xia, and V.-T. Nguyen, Adaptive fuzzy tracking control of flexible-joint robots with full-state constraints, IEEE Trans. Syst., Man, Cybern. Syst., 49(17) 2019: 2201 - 2209.
[16] G. Coletti, D. Petturiti, B. Vantaggi, Fuzzy memberships as likelihood functions in a possibilistic framework, International Journal of Approximate Reasoning, 88, January 2017: 547-566
[17] P. H. Giang, Subjective foundation of possibility theory: Anscombe-Aumann, Information Sciences, 370-371 (2016): 368-384.
[18] L. X. Wang, Fuzzy systems: Challenges and Chance--My Expreriences and Perspectives, ACTA Automatica Sinica, 27(4), 2001:585-590.
[19] F. Y. Wang, A Note on Research in Fuzzy Systems, ACTA Automatica Sinica, 28(4), 2002:663-669.
[20] H. A. P. Blom and Y. Bar-Shalom, The interacting multiple model algorithm for systems with Markov switching coefficients, IEEE Transactions on Automatic Control, vol. 33, no. 8, pp. 780-783, Aug. 1988
[21] G. Wang, X. Wang and Y. Zhang, Variational Bayesian IMM-Filter for JMSs with Unknown Noise Covariances, IEEE Transactions on Aerospace and Electronic Systems, July 2019, DOI: 10.1109/TAES.2019. 2929975.
[22] D. Dubois, and Henry Prade. Possibility Theory and Its Applications: Where Do We Stand? *Springer Handbook of Computational Intelligence*. Springer Berlin Heidelberg, 2015.
[23] B. Solaiman, É. Bossé, Possibility Theory for the Design of Information Fusion Systems, Springer, Cham, 2019.
[24] S. Lapointe, B. Bobée, Revision of possibility distributions: A Bayesian inference pattern, Fuzzy sets and systems, 116(2) (2000): 119-140.
[25] B. Ristic, J. Houssineau, S. Arulampalam, Target tracking in the framework of possibility theory: The possibilistic Bernoulli filter. Information Fusion, 2020, 62.
[26] S. Blackman, R. Popoli, Design and Analysis of Modern Tracking Systems, Artech House, Boston, January 1999
[27] W. Mei, G. L. Shan, and X. R. Li, Simultaneous tracking and classification: a modularized scheme, *IEEE Transactions on Aerospace and Electronic Systems*, 43(2) (2007): 581-599.
[28] J. Houssineau and A.N. Bishop. Smoothing and filtering with a class of outer measures. SIAM Journal on Uncertainty Quantifification, 2018.
[29] G.J. Klir, J.F. Geer, in: Fuzzy Logic, Springer Netherlands, Dordrecht, 1993, pp. 417–428, https://doi.org/10.1007/978-94-011-2014-2_39.
[30] W. Cao, J. Lan, X.R. Li, Conditional Joint Decision and Estimation With Application to Joint Tracking and Classification, IEEE Trans. Syst. Man Cybern, Syst. 46 (4) (2016) 459–471.
[31] J. Zhu, J. Chen, W. Hu, et al, Big Learning with Bayesian methods. National Science Review, 4 (2017): 627-651.
[32] W. Mei, G.-L. Shan, Y.-F. Wang, A second-order uncertainty model for target classification using kinematic data, Information Fusion 12 (2) (2011) 105–110.
[33] Ferrero, H. V. Jetti and S. Salicone, The Possibilistic Kalman Filter: Definition and Comparison with the Available Methods, in IEEE Transactions on Instrumentation and Measurement, vol. 70, pp. 1-11, 2021.
[34] X. R. Li and V. P. Jilkov, Survey of maneuvering target tracking. Part V. Multiple-model methods, *IEEE Transactions on Aerospace and Electronic Systems*, vol. 41, no. 4, pp. 1255-1321, Dec. 2005.
[35] Y. Ma, S. Zhao, and B. Huang, Multiple-model state estimation based on variational Bayesian inference, *IEEE Transactions on Automatic Control*, vol. 64, no. 4, pp. 1679-1685, Apr. 2019.
[36] S. Y. Cho, Measurement error observer-based IMM filtering for mobile node localization using WLAN RSSI measurement, *IEEE Sensors Journal*, vol. 16, no. 8, pp. 2489-2499, Apr. 2016
[37] L. B. Cosme, W. M. Caminhas, M. F. S. V. D'Angelo and R. M. Palhares, A Novel Fault-Prognostic Approach Based on Interacting Multiple Model Filters and Fuzzy Systems, IEEE Transactions on Industrial Electronics, vol. 66, no. 1, pp. 519-528, Jan. 2019.
[38] X. Feng, Y. Zhao, Z. Zhao and Z. Zhou, Cognitive Tracking Waveform Design Based on Multiple Model Interaction and Measurement Information Fusion, IEEE Access, vol. 6, pp. 30680-30690, 2018.
[39] C. B. Chang, M. Athans, State Estimation for Discrete Systems with Switching Parameters, IEEE Transactions on Aerospace and Electronic Systems 14(3) 1978:418-425
[40] J. K. Tugnait, Detection and estimation for abruptly changing systems. Automatica, 18(5) 1982: 607-615.
[41] A. Averbuch, S. Itzikowitz, T. Kapon. Radar target tracking-Viterbi versus IMM, IEEE Transactions on Aerospace & Electronic Systems, 1991, 27(3): 550-563.
[42] D. Lerro and Y. Bar-Shalom, Tracking with debiased consistent converted measurements versus EKF, IEEE Trans. on Aerospace and Electronic Systems 29 (3) (1993) 1015–1022.
[43] W. Mei and Y. Bar-Shalom, Unbiased Kalman filter using converted measurements: revisit, in: Proc. of SPIE Conference on Signal and Data Processing of Small Targets, (7445-38) (2009), San Diego, CA, USA.
[44] W. Sen, Q. Bao, Decorrelated unbiased converted measurement for bistatic radar tracking, *Journal of Applied Remote Sensing*, 15(01) 2021
[45] L. A. Johnston, and V. Krishnamurthy . An improvement to the interacting multiple model (IMM) algorithm. *IEEE Transactions on Signal Processing* 49.12(2001):2909-2923.
[46] Marco H. Terra, J. Y. Ishihara, and A. P. Junior, Array Algorithm for Filtering of Discrete-Time Markov Jump Linear Systems. *IEEE Transactions on Automatic Control* 52.7(2007):1293-1296.
[47] W. Li and Y. Jia, An information theoretic approach to interacting multiple model estimation, *IEEE Transactions on Aerospace and Electronic Systems*, vol. 51, no. 3, pp. 1811-1825, Jul. 2015
[48] E. Tsamoura, L. Michael, Neural-Symbolic Integration: A Compositional Perspective, the 35th AAAI conference of Artificial Intelligence, Feb. 2-9, 2021
[49] F. Ilhan, O. Karaahmetoglu, I. Balaban, et al, Markov RNN: An Adaptive Time Series Prediction Network with HMM-based Switching for Nonstationary Environments, arXiv preprint arXiv:2006.10119v1 (2020)
[50] H. Kohler and S. Link, Possibilistic Data Cleaning, IEEE Transactions on Knowledge and Data Engineering, doi: 10.1109/TKDE.2021.3062318.
[51] J. Houssineau, A Linear Algorithm for Multi-Target Tracking in the Context of Possibility Theory, IEEE Transactions on Signal Processing, vol. 69, pp. 2740-2751, 2021, doi: 10.1109/TSP.2021.3077304.
[52] H. Yin, J. Huang and Y. -H. Chen, Possibility-Based Robust Control for Fuzzy Mechanical Systems, IEEE Transactions on Fuzzy Systems, doi: 10.1109/TFUZZ.2020.3028940.
[53] B. Alsahwa, B. Solaiman, S. Almouahed, et al. Iterative Refinement of Possibility Distributions by Learning for Pixel-Based Classification, IEEE Transactions on Image Processing, 2016, 25(8):3533-3545.
[54] Z. He, L. Yu, Possibility distribution based lossless coding and its optimization, *Signal Processing*, Vol. 150, (2018): 122-134
[55] Z. Ren, H. Cho, J. Yeon, et al. A new reliability analysis algorithm with insufficient uncertainty data for optimal robust design of electromagnetic devices, IEEE Transactions on Magnetics, 2015, 51(3): 7207404.
[56] W. Mei, G. Li, Y. Xu, L. Shi, (2021): Probability-Possibility Transformation under Perspective of Random-fuzzy Dual Interpretation of Unknown Uncertainty. TechRxiv. Preprint. https://doi.org/10.36227/techrxiv.16896205.v1